\documentclass[journal]{IEEEtran}

\ifCLASSINFOpdf
\else
   \usepackage[dvips]{graphicx}
\fi
\usepackage{url}
\usepackage{cite}
 \usepackage{cancel}
 \usepackage{amsmath,amssymb,amsthm,graphicx,subfigure,float,stmaryrd, mathtools}
\usepackage{xcolor}
\usepackage{hyperref}
\usepackage{bm}
\usepackage{tikz}
\usepackage{enumitem}
\newcommand{\R}{\mathbb{R}}
\newcommand{\N}{\mathbb{N}}
\newcommand{\dint}[2]{\llbracket #1 ,#2\rrbracket}
\newcommand{\M}{\mathcal{M}}
\newcommand{\Reps}{\R^3_\varepsilon}
\newcommand{\Gammamat}{\bm{\Gamma}}
\newcommand{\gammavect}{\bm{\gamma}}

\newcommand{\blue}{\color{black}}
\newcommand{\zerovect}{\bm{0}}
\newcommand{\avect}{\bm{a}}

\newcommand{\rvect}{\bm{r}}
\newcommand{\xvect}{\bm{x}}
\newcommand{\yvect}{\bm{y}}

\newcommand{\tvn}[1]{\Vert #1 \Vert_\text{TV}}

\newcommand{\Ltn}[1]{\left\lVert #1 \right\rVert _2}

\newcommand{\mic}[1]{\rvect^\text{mic}_{#1}}
\newcommand{\src}[1]{\rvect^\text{src}_{#1}}

\begin{document}

\title{Gridless 3D Recovery of Image Sources from Room Impulse Responses}

\author{Tom Sprunck, \textcolor{black}{Antoine Deleforge}, Yannick Privat and  Cédric Foy 
\thanks{This work was made with the support of the French National Research Agency through project DENISE (ANR-20-CE48-0013).}
\thanks{Tom Sprunck and Antoine Deleforge are with Universit\'e de Lorraine, CNRS, Inria, LORIA, F-54000 Nancy, France (firstname.name@inria.fr).}
\thanks{Yannick Privat and Tom Sprunck are with IRMA, Universit\'e de Strasbourg, CNRS UMR 7501, 67084 Strasbourg, France (yannick.privat@unistra.fr).}
\thanks{Cédric Foy is with UMRAE, Cerema, Univ. Gustave Eiffel, Ifsttar, Strasbourg, 67035, France (cedric.foy@cerema.fr).}
\vspace{-7mm}
}

\markboth{Preprint, November 2022}
{Shell \MakeLowercase{\textit{et al.}}: Bare Demo of IEEEtran.cls for IEEE Journals}
\maketitle

\begin{abstract}
Given a sound field generated by a sparse distribution of impulse image sources, can the continuous 3D positions and amplitudes of these sources be recovered from discrete, band-limited measurements of the field at a finite set of locations, \textit{e.g.}, a multichannel room impulse response? Borrowing from recent advances in super-resolution imaging, it is shown that this non-linear, non-convex inverse problem can be efficiently relaxed into a convex linear inverse problem over the space of Radon measures in $\mathbb{R}^3$. The \textcolor{black}{new} linear operator introduced here stems from the fundamental solution of the
wave equation combined with the receivers' responses.
An adaptation of the Sliding Frank-Wolfe algorithm is proposed to numerically solve the problem \textit{off-the-grid}, \textit{i.e.}, in continuous 3D space. \textcolor{black}{Idealized} simulated experiments show that the approach \textcolor{black}{can recover hundreds of image sources at a rate and accuracy that are not achievable by previous methods, using a compact microphone array and source placed at random in random-sized shoe-box rooms}. The impact of noise, sampling rate and array diameter on these results is also examined.

\end{abstract}


\IEEEpeerreviewmaketitle

\section{Introduction}
\IEEEPARstart{W}{hen} an omnidirectional point source located at $\src{0}\in\mathbb{R}^3$ emits an impulse $\delta_0(t)$ inside an \textcolor{black}{empty enclosure}, the resulting sound pressure field $p:\mathbb{R}^3\times[0,\infty)\rightarrow\mathbb{R}$ obeys the wave equation with a source term together with boundary conditions. While this set of partial differential equations does not admit a general explicit solution, \textcolor{black}{the particular case of a rectangular room} with \textcolor{black}{rigid specular boundaries} can be treated using the celebrated \textit{image source method} of Allen and Berkley \cite{Allen1976ImageMF}, thanks to an equivalence with the following \textit{free field} inhomogeneous wave equation:
\begin{equation}
\label{eq:IS_wave}
\frac{1}{c^2}\frac{\partial^2 p}{\partial t^2} (\rvect,t) - \Delta p(\rvect,t) = \textstyle\sum_{k=0}^{+\infty} a_{k} \delta_{\src{k}}\left(\rvect\right)\delta_0(t)
\end{equation}
were $c$ denotes the speed of sound in $\textrm{m}/\textrm{s}$. Intuitively, the walls of the room have been removed and replaced by an infinite constellation of \textit{image sources} located at $\{\src{k}\}_{k\in\N^*}\subset\R^3$ that synchronously emit the same impulse $\delta_0(t)$ and correspond to iterated spatial reflections of the original source with respect to the walls. While this equivalence only strictly holds for perfectly reflective surfaces and $a_k=1$, it is commonly generalized by weighing image sources with coefficients $\{a_k\}_{k\in\N^*} \subset[0,1]$ to account for a proportion of sound energy absorbed by the walls.
If the sound field is measured in discrete time by $M$ omnidirectional microphones placed at $\{\mic{m}\}_{m\in\dint{1}{M}}\subset\R^3$ inside the room, their corresponding sampled signals can be expressed as
\begin{equation}\label{eq:discr_conv}
x_m[n] = (\kappa_m\ast p(\mic{m},\cdot))(n/f_s),\; n\in\dint{0}{N-1}
\end{equation}
where $\ast$ denotes continuous time-domain convolution, $f_s$ is the microphones' frequency of sampling in Hz, $n$ is a discrete time index and $\kappa_m:\mathbb{R}\rightarrow\mathbb{R}$ is a time-domain filter modeling the response of microphone $m$, which may also include the source response. Such signals are called \textit{room impulse responses} (RIRs). 
Convolving them with any dry source signal can be used to emulate the corresponding reverberant recorded signal.
There is an abundant literature on how to measure them in practice \cite{stan2002comparison,farina2007advancements} and a number of simulators that can compute $\xvect=(x_m[n])_{m,n}\in\mathbb{R}^{MN}$ efficiently\footnote{\textcolor{black}{Note that late reverberation models in cluttered environment should also include scattering and diffraction, but this is not the focus of this article.}} given the room dimensions, the wall reflection coefficients and the source and microphone positions, \textit{e.g.} \cite{scheibler2018pyroomacoustics}. While this \textit{forward} physical process is very well understood, fully \textit{reversing} it to recover image source parameters $\{\src{k},a_k\}_{k\in\dint{0}{K}}$ from $\xvect$ remains an open \textcolor{black}{and active} research topic, whose application domains span sound scene navigation \cite{tylka2015comparison}, auditory augmented reality \cite{neidhardt2022perceptual}, room acoustic diagnosis \cite{dilungana2022geometry}, and hearing aids \cite{kates2001room}.

\section{Related Work and Contribution}
The problem of recovering image sources from measured audio signals can be viewed as \textcolor{black}{a generalization of many tasks that have been independently investigated in the acoustic signal processing literature over the past decade}. Estimating the absolute or relative \textit{times of arrival} of image sources at microphones, also known as early \textit{echoes}, is the focus of \cite{kowalczyk2013blind, crocco2016estimation, di2020blaster, shlomo2021blindloc} and can be of independent interest in the context of \textit{echo-aware} signal processing, as reviewed in \cite{carlo2021dechorate}. Localizing \textit{reflectors} in the room is equivalent to localizing their corresponding first-order image source together with the true source. Localizing all external reflectors is popularly known as \textit{hearing the shape of the room}. Most studies on this first estimate echoes and/or directions of arrival of image sources, then label and sort them, and finish by triangulation \cite{tervo2010estimation,antonacci2012inference,sun2012localization,mabande2013room,dokmanic2013acoustic,jager2016room,remaggi2016acoustic,el20173d,lovedee2019three}. Alternatively, \cite{ribeiro2011geometrically} proposes a more direct approach based on sparse optimization. Retrieving the coefficients $a_k$ associated to reflectors in frequency bands is studied in \cite{shlomo2021blind} and \cite{dilungana2022geometry}, as they relate to their acoustic \textit{impedance}. Finally, recovering image sources within a given range is the focus of recent non-parametric sound-field reconstruction methods \cite{koyama2019sparse, damiano2021soundfield}. All these tasks can either be performed using RIRs as in \cite{ribeiro2011geometrically,antonacci2012inference,dokmanic2013acoustic,kowalczyk2013blind,crocco2016estimation,jager2016room,remaggi2016acoustic,el20173d,remaggi2018acoustic,lovedee2019three,dilungana2022geometry} or \textit{blindly} using unknown source signals as in \cite{tervo2010estimation,antonacci2012inference,sun2012localization,mabande2013room,koyama2019sparse,di2020blaster,damiano2021soundfield,shlomo2021blindloc,shlomo2021blind}.

While the above referenced studies developed a rich variety of methodologies, nearly all of them have in common the definition of a \textit{discrete grid} in 1D time \cite{tervo2010estimation,sun2012localization,mabande2013room,antonacci2012inference, dokmanic2013acoustic,kowalczyk2013blind, crocco2016estimation, jager2016room,remaggi2016acoustic,el20173d,lovedee2019three,shlomo2021blindloc,dilungana2022geometry}, in 2D space \cite{sun2012localization,mabande2013room,remaggi2018acoustic,koyama2019sparse,damiano2021soundfield} or in 3D space \cite{ribeiro2011geometrically}, as well as the use of sparse optimization techniques and/or peak-picking techniques over such grids. This \textit{on-the-grid} paradigm suffers from intrinsic limitations. First, in 3D, the required grid size grows cubically in the desired range and precision. \textcolor{black}{This fundamentally limits the accuracy of current sparse methods under reasonable computational constraints \cite{ribeiro2011geometrically,koyama2019sparse,damiano2021soundfield}}. Second, \textcolor{black}{time-domain peak-picking fails} when peaks are overlapping and distorted due to filtering effects such as $\kappa_m$. \textcolor{black}{Existing methods address this by using ad-hoc source and microphone placements inside the room \cite{antonacci2012inference,dokmanic2013acoustic,jager2016room,remaggi2016acoustic,el20173d,lovedee2019three}}. Third, sparse optimization over a discrete grid fundamentally suffers from the so-called \textit{basis-mismatch} problem \cite{chi2011sensitivity,denoyelle2019sliding}, requiring the use of ad-hoc post-processing steps.

In parallel, recent theoretical and methodological advances on the general problem of recovering spikes \textit{off-the-grid} have emerged \cite{de2012exact,candes2014towards,duval2015exact, morgenshtern2016super,denoyelle2019sliding,traonmilin2020basins,benard2022fast}, notably motivated by applications to \textit{super-resolution} in, \textit{e.g.}, fluorescence microscopy \cite{huang2009super}. Except for a couple of recent studies on blind echo estimation \cite{di2020blaster} and anechoic beamforming \cite{chardon2021gridless}, these advances seem not to have received significant attention from the audio and acoustics communities as of yet. In this article, a connection to this field is established by showing that the non-linear, non-convex inverse problem of recovering $\{\src{k},a_k\}_{k\in\dint{0}{K}}$ from $\xvect$ can be relaxed into a convex linear inverse problem over the infinite-dimensional space of Radon measures in $\mathbb{R}^3$. While existing super-resolution applications typically consider Fourier or Laplace operators, a new linear operator derived from the acoustic measurement model \eqref{eq:IS_wave} and \eqref{eq:discr_conv} is introduced here. An adaptation of the sliding Frank-Wolfe algorithm \cite{denoyelle2019sliding} to this setting is shown to efficiently solve the problem numerically in continuous 3D space. \textcolor{black}{Under idealized simulated conditions, the method achieves near-exact recovery of hundreds of image sources using one compact microphone array and one source placed at random in random-sized rooms. To the best of the authors' knowledge, this is not achievable by any previously known methodology. Robustness to noise, array size and sampling rate is also examined.}


\section{Observation Model and Inverse Problem}
Equation (\ref{eq:IS_wave}) can be further generalized to an arbitrary \textit{source mass distribution} $\psi$, yielding
\begin{equation}
\label{eq:IS_wave_gen}
\frac{1}{c^2}\frac{\partial^2}{\partial t^2} p(\rvect,t) - \Delta p(\rvect,t) = \psi(\rvect)\delta_0(t),
\end{equation}
where $\psi$ belongs to the space $\M(\R^3)$ of \textit{Radon measures}, \textit{i.e.}, the topological dual of the space of continuous functions on $\mathbb{R}^3$ that vanish at
infinity \cite{denoyelle2019sliding}.
Using the linearity of the wave equation, the general solution of (\ref{eq:IS_wave_gen}) is then given by the following spatial convolution product with a Green function:
%
\begin{equation}
p(\rvect,t) = (G(\cdot,t)\ast\psi)(\rvect)
= \int_{\rvect'\in\mathbb{R}^3}\hspace{-5mm}\frac{\delta(t-\|\rvect-\rvect'\|_2/c)}{4\pi\|\rvect-\rvect'\|_2}\psi(\rvect')d\rvect'. \label{eq:solution_psi}
\end{equation}
Intuitively, $G:\mathbb{R}^3\times\mathbb{R}\rightarrow\mathbb{R}$ is a spherical wave centered at the origin propagating outwards at the speed of sound. It corresponds to the fundamental solution of the wave equation, obtained by setting $\psi(\rvect)=\delta_{\zerovect}(\rvect)$ in \eqref{eq:IS_wave_gen}. Combining \eqref{eq:solution_psi} and \eqref{eq:discr_conv}, the following expression is obtained for the discrete signal measured by a microphone $m$ placed at $\mic{m}$ observing $p$:
\begin{align}
\label{eq:gamma}
&x_m[n] =  \int_{\rvect\in\mathbb{R}^3}\gamma_{m,n}(\rvect)d\psi(\rvect)=\langle\gamma_{m,n},\psi\rangle \\
&\textrm{where}\;\; 
\gamma_{m,n}(\rvect)\stackrel{\textrm{def}}{=} \frac{ \kappa\left(n/f_s-\left\|\mic{m}-\rvect\right\|_2/c\right)}{4\pi\left\|\mic{m}-\rvect\right\|_2}.
\end{align}
Observe that the \textit{non-linear}, \textit{non-convex} function $\gammavect:\mathbb{R}^3\rightarrow\mathbb{R}^{MN}$ can be seen as the representative of an infinite-dimensional \textit{linear} operator\footnote{Strictly speaking, this operator is not well defined because $\gammavect$ is \textit{singular} at each microphone position. In theory, one should change the integration domain in \eqref{eq:gamma} to $\Reps \stackrel{\textrm{def}}{=}\R^3\setminus \cup_{m=1}^M B(\mic{m},\varepsilon)$
for a fixed $\varepsilon>0$, and only consider measures $\psi\in\M(\Reps)$. In practice, this adjustment is harmless as long as a minimum separation distance $\varepsilon$ is assumed between the image sources and the microphones, and is hence ignored here for clarity.} $\Gammamat:\M(\R^3)\rightarrow\R^{MN}$ that maps an arbitrary source mass distribution $\psi$ to its corresponding observation vector $\xvect$.
We can now particularize \eqref{eq:solution_psi} and \eqref{eq:gamma} to a \textrm{discrete} image source distribution $\psi_{\avect,\src{}}$ as in \eqref{eq:IS_wave}, yielding:
\begin{align}
\psi_{\avect,\src{}}(\rvect) &\stackrel{\textrm{def}}{=} \textstyle\sum_{k=0}^{K} a_{k}  \delta_{\src{k}}\left(\rvect\right), \\
p(\rvect,t)&=\textstyle\sum_{k=0}^{K} a_{k} \frac{ {\delta}\left(t-\left\|\rvect-\rvect_{k}^{\textrm{src}}\right\|_2/c\right)}{4\pi\left\|\rvect-\rvect_{k}^{\textrm{src}}\right\|_2}, \label{eq:solution_IS} \\
x_m[n] &= \textstyle\sum_{k=0}^Ka_k\frac{ \kappa_m\left(n/f_s-\left\|\mic{m}-\src{k}\right\|_2/c\right)}{4\pi\left\|\mic{m}-\src{k}\right\|_2}, \label{eq:RIR_IS}  \\
\xvect &= \textstyle\sum_{k=0}^{K} a_{k} \gammavect(\rvect_{k}^{\textrm{src}}) = \Gammamat\psi_{\avect,\src{}} \in\mathbb{R}^{MN}
\end{align}
where $K\in\mathbb{N}\cup\infty$. 
Notice that the RIR signals in \eqref{eq:RIR_IS} are weighted sums of delayed filters, which is how most image-source simulators are implemented in practice, \textit{e.g.}, \cite{scheibler2018pyroomacoustics}. As the image source order increases, their distances and corresponding times of arrival at microphones increase while their weights decrease. The sound pressure field $p$ and the observations $\xvect$ are hence reasonably approximated by considering a \textit{finite} value for $K$.
Let $\M_*(\R^3)\subset \M(\R^3)$ denote the subset of sparse Radon measures of the form $\psi_{\avect,\rvect^{\textrm{src}}}$ with $K\in\N$, $\avect\in\mathbb{R}_+^{K+1}$ and $\rvect^{\textrm{src}}\in(\mathbb{R}^{3})^{K+1}$, \textit{i.e.} measures that are finite positive combinations of spikes. The inverse problem of recovering the amplitudes and positions of these spikes given noisy observations $\xvect$ can be formulated as {\blue follows}:
\begin{equation}\label{eq:ls_opt_pb}
\operatorname*{argmin}_{\psi_{\avect,\src{}}\in\M_*(\R^3)}\left\|\xvect - \Gammamat\psi_{\avect,\src{}}\right\|^2_2.
\end{equation}

This belongs to a general class of problems where the goal is to recover the continuous locations of a set of spikes given discrete linear observations over their measure \cite{de2012exact,candes2014towards, morgenshtern2016super, duval2015exact, denoyelle2019sliding, traonmilin2020basins,benard2022fast}.
Rather than solving \eqref{eq:ls_opt_pb}, which is a non-convex optimization problem on the amplitudes and positions of the image sources, we follow the approach in \cite{denoyelle2019sliding} and consider a convex relaxation to the whole space of Radon measures:
\begin{equation}
\label{eq:blasso}
\operatorname*{argmin}_{\psi \in \M(\R^3)}\frac{1}{2}\Ltn{\xvect-\Gammamat\psi}^2 + \lambda\tvn{\psi}
\end{equation}
where $\lambda\in\R^*_+$ is a parameter and $\tvn{\psi}$ denotes the \textit{total variation} norm of the Radon measure $\psi$. For a sparse measure $\psi_{\avect,\rvect^{\textrm{src}}}\in\M_*(\R^3)$ we have $   \tvn{\psi_{\avect,\rvect^{\textrm{src}}}} = \textstyle\sum_{k=0}^K|a_k|=\|\avect\|_1.$
Hence, the second term can be seen as a sparsity-inducing regularizer.
By analogy with the finite-dimensional sparse setting, this problem has been coined the Beurling-LASSO (BLASSO) in \cite{de2012exact} and offers good measure-reconstruction guarantees in low noise regimes. In particular, \cite{boyer2019representer} shows that there always exists a sparse solution $\psi_{\avect,\src{}}\in\M_*(\R^3)$ to problem \eqref{eq:blasso} and \cite{traonmilin2020basins} studies its basins of attraction.


\section{Algorithm}
\begin{figure}
    \centering
    \includegraphics[scale=0.55]{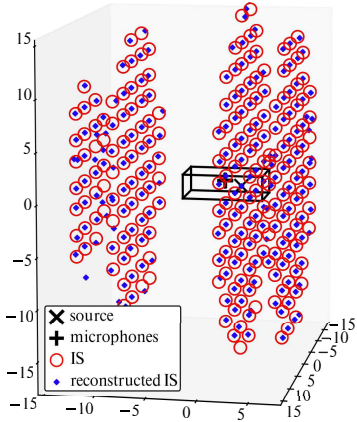}
    \caption{3D plot of a room and the corresponding target and reconstructed image sources for a 32-channel spherical microphone array with diameter 16.8 cm, $f_s = $16 kHz, $T_{\textrm{max}} = $ 50 ms and no noise. The corresponding RIR at one microphone is shown in Fig.~\ref{fig:rir}.}
    \label{fig:room}
    \vspace{-4mm}
\end{figure}

%
In order to solve \eqref{eq:blasso} numerically, we adapt the Sliding Frank-Wolfe algorithm proposed in \cite{denoyelle2019sliding}, which is briefly reviewed below.
%
Let us denote by $\avect^{(i)}\in\mathbb{R}_+^{Q_i}$ and $\rvect^{(i)}\in(\mathbb{R}^{3})^{Q_i}$ the lists of $Q_i$ spike amplitudes and positions estimated at iteration $i$, with $\avect^{(0)}=\rvect^{(0)}=\emptyset$. At iteration $i+1$, the following four steps are performed:
\begin{itemize}[leftmargin=*]
    \item \textbf{Step 1:} A new spike location $\rvect_{Q_i+1}$ is first added to $\rvect^{(i+1)}$ by maximizing the following \textit{dual}, non-convex objective based on the current residual $\yvect^{(i)}\stackrel{\textrm{def}}{=}\xvect - \Gammamat \psi_{\avect^{(i)},\rvect^{(i)}}$:
\begin{equation}
    \label{eq:opt_etak}
    \hspace{-2mm}\operatorname*{max}_{\rvect\in\R^3}\;\eta^{(i)}(\rvect) \stackrel{\textrm{def}}{=}  \left[\Gammamat^*(\yvect^{(i)})\right](\rvect) = \sum_{m,n}y^{(i)}_m[n]\gamma_{m,n}(\rvect)
\end{equation}
where $\Gammamat^*$ denotes the \textit{Hermitian adjoint} of $\Gammamat$.
%
%
%

\item \textbf{Step 2:}  The whole list of amplitudes $\avect^{(i+1)}$ is updated by minimizing \eqref{eq:blasso} over $\avect$ only. This amounts to a classical non-negative LASSO convex optimization problem for which efficient solvers are available, \textit{e.g.}, in \cite{pedregosa2011scikit}.

\item \textbf{Step 3} (\textit{Sliding}):  The value of the cost function in \eqref{eq:blasso} is further decreased by jointly refining all the values in $\avect^{(i+1)}$ and $\rvect^{(i+1)}$ through non-convex local search.

\item \textbf{Step 4:} The spikes whose amplitudes are lower than a threshold $\alpha_{\textrm{min}}$ are removed from $\avect^{(i+1)}$ and $\rvect^{(i+1)}$.
\end{itemize}

A number of modifications are introduced to improve the optimization and reduce the computational time. As in \cite{denoyelle2019sliding}, the non-convex maximization in step 1 is carried out using the BFGS algorithm, and is very sensitive to the choice of an initial guess. We use the {\blue Scipy} implementation
and propose to initialize it by first solving the problem over a discrete spatial grid. Because the mass of $\eta^{(i)}$ is tightly contained around the target image sources, a fine grid is required. Covering the entire 3D search region with such a grid would be intractable (over 2 million points). To restrict this region, a moving average over 3 samples is applied to the squared residual signal of each microphone, and the sample $\hat{n}_m$ maximizing these signals for each $m$ is calculated. At least one image source is expected to be located on the spheres $\{S(\mic{m}, cf_s\hat{n}_m)\}_m$. Hence, uniform grids with a mean angular spacing of $5^{\circ}$ are built on the spheres corresponding to the 8 microphones with highest peaks, as well as on their neighboring spheres with radii $\pm5$~cm. The initial guess is then the value maximizing $\eta^{(i)}(\rvect)$ over the union of these grids ($\sim$40k points). Note that due to the singularity of $\gammavect$ mentioned in the previous section, problem \eqref{eq:opt_etak} does not in fact admit a solution. However, experiments showed that initializing far enough from the microphone array removed this issue in practice. Iterations are stopped when the amplitude of the last recovered spike is below $\alpha_\textrm{min}=0.01$. To further improve the optimization and reduce computational time, the algorithm is first ran on an early-cut version of the time signals, where echoes are better separated. The resulting spikes are then used as an initialization to run the algorithm on progressively longer signals, and this is repeated until the desired signal length $N$ is reached. Finally, the sliding step 3 is only performed \textit{on the very last iteration}, as suggested in \cite{benard2022fast}, using {\blue the parallel} bounded BFGS {\blue implementation in \cite{gerber2020optim}} to preserve positive amplitudes. 
Spikes with an amplitude less than $0.1$ are deleted before and after sliding to decrease the number of false positives. $\lambda$ is fixed to $3\cdot 10^{-5}$ throughout the experiments based on a preliminary manual tuning. Our code for this algorithm is available at \urlstyle{tt}\url{https://github.com/Sprunckt/acoustic-sfw}.

 \begin{figure}
    \centering
    \includegraphics[scale=0.6]{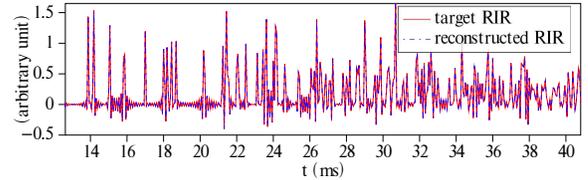}
    \caption{Excerpt from a room impulse response (RIR) at one microphone and its reconstruction using \eqref{eq:RIR_IS}, under the setup described in Fig.~\ref{fig:room}.}
    \label{fig:rir}
    \vspace{-4mm}
\end{figure}

\section{Numerical results}


We present here some of the numerical results obtained by applying the algorithm described in the previous section to a set of 200 simulated rooms containing an omnidirectional source and a spherical array of 32 microphones. The geometry of the array is the same as the em32 Eigenmike\textsuperscript{\textregistered} (diameter $d$=8.4~cm) but scaled by various factors \textcolor{black}{and using an open-sphere model}. We use a unique ideal low-pass filter with cutoff frequency $f_s/2$ to model the response of all the microphones, \textit{i.e.}, $\kappa_m(t) = \operatorname{sinc}(\pi f_s t)$ for all $m$. The rooms' lengths and widths in meters are sampled uniformly at random in $[2,10]$, while the heights are taken in $[2,5]$. The absorption coefficient of each individual wall is sampled uniformly at random in $[0.01,0.3]$. The source and the array are then placed randomly in each room, with a separation constraint of $1$ m to the walls and between each other\footnote{\textcolor{black}{This unique constraint was chosen for simplicity, but could be dropped between mic. and walls and safely relaxed between sources and walls.}}.
\textcolor{black}{While full-lenght RIRs are simulated, the proposed approach is only fed with the first $T_{\textrm{max}}=(N-1)/f_s=50$~ms of each channel. This allows us to consider as targets all the image sources that are \textit{audible} by all the microphones, \textit{i.e.}, whose distances are inferior to $cT_{\textrm{max}}=17.15$~m, independently of the room dimension}.
For each test room, the ground truth image source positions and amplitudes are obtained using the pyroomacoustics simulator \cite{scheibler2018pyroomacoustics} with $c=343 \textrm{m}/\textrm{s}$. An observation vector is then built using \eqref{eq:RIR_IS} and adding white Gaussian noise with a desired peak signal-to-noise ratio (PSNR). An example of room, image source constellation, RIR, recovered image sources and reconstructed RIR is shown in Fig.~\ref{fig:room} and Fig.~\ref{fig:rir}. 

\begin{table}
    \centering
        \caption{Mean room volume ($\overline{\textrm{V}}$), Recall (R), Precision (P) and mean radial ($\overline{\textrm{RE}}$), angular ($\overline{\textrm{AE}}$), Euclidian ($\overline{\textrm{EE}}$) and amplitude ($\overline{\textrm{AmE}}$) errors amongst the recovered sources for varying numbers of image sources, with $f_s$=16~kHz, $d$=16.8~cm and no noise.}
         \setlength{\tabcolsep}{4.pt}
    \begin{tabular}{c|c||c|c|c|c|c|c}
    \# of IS & $\overline{\text{V}}$($\text{m}^3$) &
    R(\%) & P(\%) &
    $\overline{\text{RE}}$(mm) & $\overline{\text{AE}}$(°) & $\overline{\text{EE}}$(mm) & $\overline{\textrm{AmE}}$ \\\hline 
0-150    & 214 & 94.3 & 81.8 & 0.069 & 0.38 & 94  & 0.042\\
150-300  & 102 & 92.1 & 83.1 & 0.099 & 0.36 & 91  & 0.029\\
300-500  & 56  & 86.1 & 78.1 & 0.151  & 0.38 & 97  & 0.025\\
500-1323 & 30  & 57.3 & 51.6 & 0.300  & 0.46 & 108 & 0.027\\
    \end{tabular}
    \label{tab:metrics}
    \vspace{-4mm}
\end{table}

To evaluate the efficiency of the method, a source is considered \textit{recovered} if at least one estimated source is at an angular distance of less than $2^{\circ}$ and a radial distance of less than $1$~cm from it with respect to the array center. We then calculate the recall (ratio of true sources recovered), the precision (ratio of estimated sources assigned to a recovered source, discarding doubles), as well as the mean radial, angular and Euclidean errors and the mean error on amplitudes, where the means are calculated \textit{over recovered sources only}. Because the number of image sources that are audible within 50~ms of RIRs varies widely depending on the room's volume, the test set is sliced into four subsets, as detailed in the first two columns of Table~\ref{tab:metrics}. The remaining columns report the metrics for a sampling frequency $f_s$=16~kHz, an array diameter $d$=16.8~cm (x2) and no noise. A recall rate of over $90\%$ for large and medium sized rooms is obtained. The precision is over $80\%$, indicating few false positives and a reasonable prediction of the number of audible sources. As expected, the recall and precision significantly drop in smaller rooms, where the \textit{echo density} \cite{tukuljac2019sparsity} is higher, making the image sources harder to separate. The strength of the proposed gridless approach is revealed by the mean radial and angular errors, which are below tenths of millimeters and fractions of degrees. \textcolor{black}{As a first comparison, the best previously reported results we are aware of in a similar simulated setting are in \cite{shlomo2021blind}, where an average of 25 nearest image sources are localized with a mean angular error of $4.3^\circ$. Note however that \cite{shlomo2021blind} is a \textit{blind} method. As a second comparison, ignoring any basis-mismatch issue and assuming \textit{perfect localization}, a sparse method in discrete space such as \cite{ribeiro2011geometrically} would require a spatial grid of at least 111 million points to achieve errors below $1^\circ$ and 1~cm over the same range. This is four orders of magnitude larger than the grids used in the proposed approach.}
%
%

Notice that the obtained mean Euclidean errors are of a few centimeters. This is because they grow with the source distances, as expected due to the compact spherical geometry of the array. The amplitudes of the recovered sources are also accurately estimated, with mean errors around 0.03 (note that amplitudes lie in $[0,1]$). These errors are slightly larger in large rooms because amplitudes are larger in that case, due to fewer reflections on the walls. Note that only 3 out of 1200 first-order image sources were missed on this test, while all 200 true sources were recovered.

\begin{figure}[!t]
    \centering
    \includegraphics[scale=0.75]{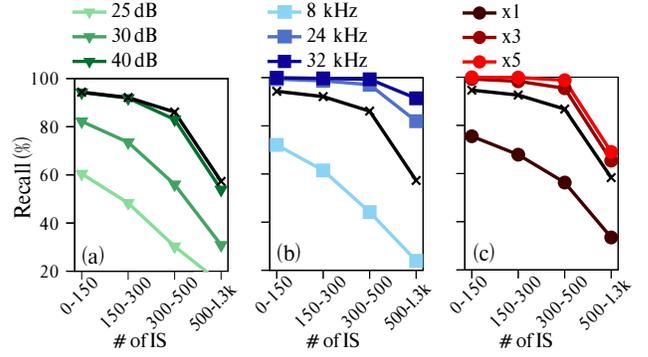}
    \caption{Recall for varying PSNR (a), sampling frequency (b) and microphone array scaling (c). The default values (\textbf{\texttimes}) are noiseless, 16~kHz and x2.}
    \label{fig:recall}
    \vspace{-4mm}
\end{figure}

The impact of PSNR, $f_s$ and $d$ on the recall is reported in 
Fig. \ref{fig:recall}. Remarkably, it can be observed that either increasing $f_s$ to 32~kHz or the array diameter to 42~cm (x5) brings the recovery rate near 100\% for rooms with up to 500 image sources. Conversely, decreasing by half these parameters significantly degrades performance. This is expected as they are known to control the source localization accuracy for compact microphone arrays.
Adding noise to the observations does not significantly affect the recovery rate at 40~dB PSNR, but quickly degrades it for PSNRs below 30~dB. Nevertheless, it was observed that the recall values for a PSNR of 30~dB could be restored near the noiseless level by simply considering an angular recovery threshold of $6^{\circ}$ instead of $2^{\circ}$. This shows an encouraging stability of the method, given that for such PSNRs the peaks of many echoes in the RIRs fell below the noise standard deviation.

\section{Conclusion and Future Work}
We introduced a new method to recover the continuous 3D positions and amplitudes of all audible image sources given the early part of a discrete-time multichannel RIR from a compact microphone array. \textcolor{black}{While the obtained recovery results under idealized conditions are unprecedented to the best of our knowledge, applying the method to real data will require a number of challenging extensions. Indeed, real RIRs are impacted by the frequency and angular dependencies of source, microphone and wall responses. This will require new formulations of problem \eqref{eq:ls_opt_pb} in the Fourier and spherical-harmonic domains and an extension of the framework to frequency-dependent amplitudes.
Generalization to non-cuboid polyhedral rooms will require robust extensions to \textit{occlusions}, \textit{i.e.}, image sources that are only audible by a subset of microphones.
To improve robustness, extending the approach to multiple source and receiver placements in the room is a worthwhile direction.
Finally, theoretical investigations on the solutions to problem \eqref{eq:blasso} as well as applications to reflective surface localization and analysis will be pursued.} 

\bibliographystyle{unsrt}
\bibliography{biblio} 

\end{document}